 \documentclass[final,5p,times,twocolumn]{elsarticle}

\usepackage{amssymb}
\usepackage{amsmath}
\usepackage{xcolor}
\usepackage[utf8]{inputenc}

\usepackage{xr}

\makeatletter
\newcommand*{\addFileDependency}[1]{
\typeout{(#1)}
\@addtofilelist{#1}
\IfFileExists{#1}{}{\typeout{No file #1.}}
}\makeatother

\DeclareUnicodeCharacter{2212}{\textminus}

\journal{Surface Science -- Special Issue: 60 years of Surface Science: Achievements and Perspectives}

\begin{document}

\begin{frontmatter}



\title{Pyridyl-functionalized tripod molecules on Au(111): Interplay between H-bonding and metal coordination}


\author[first]{Sajjan Mohammad\fnref{fn1}}
\author[second]{Neeta Bisht\fnref{fn1}}
\author[first]{Anjana Kannan}
\author[second]{Anne Brandmeier}
\author[second]{Christian Neiss}
\author[second,third]{Andreas Görling\corref{cor1}}
\ead{andreas.goerling@fau.de}
\author[first,fourth]{Meike Stöhr\corref{cor1}}
\ead{meike.stoehr@fau.de}
\author[first]{Sabine Maier\corref{cor1}}
\ead{sabine.maier@fau.de}

\affiliation[first]{organization={Department of Physics, Friedrich-Alexander-Universität Erlangen-Nürnberg},
            addressline={Erwin-Rommel-Str.~1}, 
            city={Erlangen},
            postcode={91058}, 
            country={Germany}}
\affiliation[second]{organization={Department of Chemistry and Pharmacy, Chair of Theoretical Chemistry, Friedrich-Alexander-Universität Erlangen-Nürnberg},
            addressline={Egerlandstr.~3}, 
            city={Erlangen},
            postcode={91058},    
            country={Germany}}
\affiliation[third]{organization={Erlangen National High Performance Computing Center (NHR@FAU)},
            addressline={Martensstr. 1}, 
            city={Erlangen},
            postcode={91058}, 
            country={Germany}}            
\affiliation[fourth]{organization={University of Applied Sciences of the Grisons},
            addressline={Pulvermühlestrasse~57}, 
            city={Chur},
            postcode={7000}, 
            country={Switzerland}}
            
\cortext[cor1]{Corresponding author}
\fntext[fn1]{These authors contributed equally.}

\begin{abstract}
The self-assembly of pyridyl-functionalized triazine (T4PT) was studied on Au(111) using low-temperature scanning tunneling microscopy (STM) under ultra-high vacuum conditions combined with density functional theory (DFT) calculations. In particular, we investigated the effect of temperature on the intermolecular interactions within the assemblies. STM measurements revealed that T4PT molecules form a well-ordered, close-packed structure, with the molecules adopting a planar conformation parallel to the Au surface for coverages $\leq1$~ monolayer upon room temperature deposition. The intermolecular interactions stabilizing the self-assembled arrangement is based on a combination of hydrogen bonding and weak van der Waals forces. Upon post-deposition annealing, the assemblies are additionally stabilized by metal-ligand bonding between the pyridyl ligands and native Au adatoms. Further post-deposition annealing at temperatures above $200^{\circ}$~C led to the breaking of the N-Au bonds with the molecular assemblies transforming into a second close-packed hydrogen bonded structure. For temperatures exceeding $230^{\circ}$~C, few covalently linked dimers formed, most likely as a result of CH-bond activation. We rationalize the kinetically-driven structure formation by unveiling the interaction strengths of the bonding motifs using DFT and compare the molecular conformation to the structurally similar pyridyl-functionalized benzene (T4PB).
\end{abstract}




\end{frontmatter}




\section{Introduction}
\label{introduction}


Two-dimensional metal-organic coordination networks (MOCNs) on metal surfaces represent an intensely researched material class, as the tunability and adaptability of coordination bonding offer rich opportunities for designing and fabricating novel materials with potential applications in fields such as catalysis and molecular electronics.\cite{Dong2016review,Liu2023review,Zhao2024review} The metal-ligand interaction offers the advantage of a relatively high stability for a supramolecular structure. A strategic choice of metal ions and organic ligands can produce highly ordered functional structures, allowing for precise control of their structural, electronic, and magnetic characteristics. For example, theoretical calculations have identified 2D MOCNs as potential organic topological insulators \cite{Wang2013} and they are candidates to exhibit two-dimensional ferromagnetism at the single-layer limit.\cite{Umbach2012, Abdurakhmanova2013}

In particular, MOCNs utilizing pyridyl groups as coordination motifs have garnered attention, as the pyridyl groups can be involved with their nitrogen atoms in both hydrogen-bonding interactions  (the N atoms act then as hydrogen-bond acceptors) and metal-ligand bonding, typically with metal (ad)atoms. Due to the relatively high bonding strength as well as good tunability of the coordination motif in dependence of the used metal, pyridyl-functionalized molecules have often been the choice for fabrication of long-range ordered porous as well as close-packed 2D assemblies on surfaces.\cite{Studener2015,Heim2010,Liujun2011,Vijayaraghavan2013,Shi2009,Song2017_py,Zhao2017,Piquero2019,WangWeihua2013,LiBo2022,LiYang2012,Adisoejoso2012} Over the past two decades, the usage of atoms supplied from the metal substrate (so-called native atoms), onto which the molecules are adsorbed, has become a popular alternative to additionally supplying metal atoms from an external source in order to establish metal-ligand bonding. This approach has often been used on Cu(111).\cite{Klappenberger2008,Jonas2010}  For MOCNs on Au(111), only a very few examples have been so far presented utilizing the Au-pyridyl coordination bonding for stabilizing the 2D arrangement.\cite{Shi2009,Song2017_py,Lu2022,Geagea2021} In contrast, mostly co-deposited Fe or Co atoms have been used for the fabrication of MOCNs, which bind stronger to pyridyl groups.\cite{DeYin2002}

\begin{figure}[ht]
	\centering 
	\includegraphics[width=0.4\textwidth]{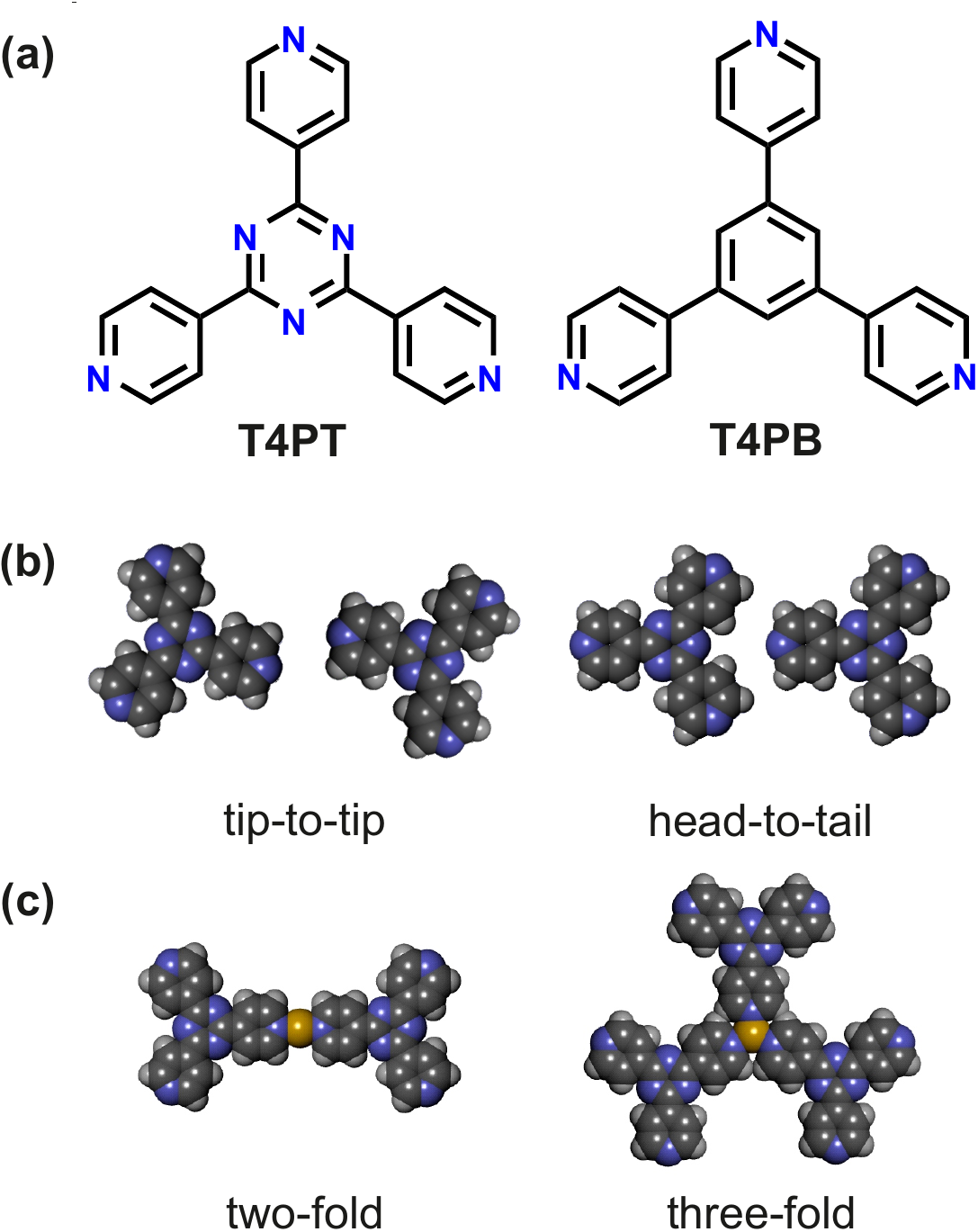}	
	\caption{(a) Molecular structure of 2,4,6-tri(4-pyridyl)-1,3,5-triazin (T4PT) and 1,3,5-tri(4-pyridyl)benzene (T4PB). (b) Possible H-bonding and (c) metal-coordination bonding motifs for T4PT. C, N, H, and Au atoms are depicted in gray, blue, white, and yellow, respectively. (For interpretation of the references to color in this figure legend, the reader is referred to the web version of this article.)} 
	\label{fig_scheme}%
\end{figure}

In this article, we report on the self-assembly of 2,4,6-tri(4-pyridyl)-1,3,5-triazine (T4PT) and 1,3,5-tri(4-pyridyl)benzene (T4PB) on Au(111) using low-temperature scanning tunneling microscopy (STM) under ultra-high vacuum (UHV) conditions combined with density functional theory (DFT) calculations. T4PT and T4PB consist of three pyridyl groups connected to a central triazine and benzene ring, respectively (see Fig.~\ref{fig_scheme}). The on-surface self-assembly of T4TP \textit{via} hydrogen-bonding \cite{Zhu2022,Lu2022} as well as metal-ligand bonding \cite{Umbach2012,Umbach2014,Zhou2020T4PT,Lu2023,Lu2022}  has been previously studied on coinage metal surfaces using STM in UHV. Here, we discuss different self-organized assemblies of T4PT based on hydrogen bonding, metal-ligand bonding with native Au adatoms, and van der Waals (vdW) interactions on Au(111), whereas not all three interaction types are present for each arrangement.  Our results demonstrate that the various occurring intermolecular interactions compete with one another, hampering predictability of and control over the 2D arrangement of pyridyl derivatives on metal surfaces.


\section{Experimental method}
\label{methods}

\noindent \textbf{STM}. The STM measurements were carried out at T = 4.2~K using a commercial STM/nc-AFM (Scienta-Omicron GmbH) equipped with a Nanonis control unit (SPECS GmbH) in a UHV system with a base pressure lower than 3$\times 10^{-10}$~mbar. A qPlus tuning fork sensor ($k\approx 1800$~Nm$^{-1}$) with a chemically etched tungsten tip was used to acquire the STM images.\cite{Giessibl03} The STM tips were prepared and formed by controlled indentation into the Au(111) surface. The tip was grounded, and the bias was applied to the sample during the measurements. All bias voltages mentioned in the manuscript refer to the sample bias. The STM data were processed using the WSxM software package.\cite{Horcas07}  \\

\noindent \textbf{Sample preparation}. The Au(111) single crystal (MaTeck) was cleaned by several cycles of argon ion sputtering followed by annealing at 690~K for 15~min. All sample temperature values were obtained with a thermocouple located in close proximity to the sample. Commercially available 2,4,6-Tri(4-pyridyl)-1,3,5-triazine (TCI, purified by sublimation, $>97.0$\%) and 1,3,5-tri(4-pyridyl)benzene (BLD Pharmatech GmbH, $97.0$\%) was thermally evaporated in UHV from a Knudsen cell (Kentax GmbH) located in the preparation chamber. The sublimation from a quartz crucible occurred at 423~K. During deposition, the Au(111) sample was kept at 300~K if not otherwise indicated in the figure caption. The deposition rate was calibrated by a quartz microbalance and the powders were thoroughly degassed before sublimation onto Au(111). The deposition rate used was around 0.1-0.2 ML/min; a monolayer refers to the coverage of the close-packed assembly presented in Fig.~\ref{fig_STM_RT}. Post-deposition annealing of the samples was performed through resistive heating.\\

\noindent \textbf{DFT calculations}. DFT calculations were performed using the Vienna Ab initio Simulation Package (VASP) in the projector augmented wave (PAW) framework, where the valence electrons are described by a plane wave basis set and the atomic cores by the PAW method.\cite{Kresse1996,Kresse1996b,Kresse1999} The generalized gradient approximation (GGA) of Perdew–Burke–Ernzerhof (PBE) for the exchange-correlation functional was employed.\cite{Perdew96}  To account for van der Waals forces, we used the DFT+D3 method with Becke-Johnson damping function.\cite{Grimme10}. The input structures for the unit cells consisted of a preoptimized Au(111) slab with four layers of gold atoms. Three layers were fixed to the bulk values of a calculated Au(111) unit cell. Only the uppermost layer was allowed to relax structurally. Between the slabs, there was 15~\AA \ of vacuum such that the interaction between the slabs is negligible. We consider an orthorhombic unit cell for Au(111) just for calculation purposes. 
A $\gamma$-centered $k$-point sampling has been used to sample the Brillouin zone during geometry optimizations. The energy cut-off for the plane wave is kept at 400~eV for the ground state calculations. Energies were converged to 10$^{-6}$~eV and geometries were relaxed until the forces on all atoms were below 
0.01~eV/\AA, respectively. In all cases, a Gaussian smearing with $\sigma$=0.01~eV was used.\\ 

\section{Results}

\subsection{Bonding motifs}
Before we turn to the results for adsorption of T4PT on Au(111), we will discuss in the following possibly occurring bonding motifs between the molecules which are summarized in Fig~\ref{fig_scheme}b and c. In the gas phase, T4PT adopts a fully planar conformation due to hydrogen bonding between the H atoms of its pyridyl units and the N atoms of the central triazine core.  This conformation was obtained through energy minimization using DFT at the PBE+D3 level in the gas phase, see Fig.~\ref{fig_T4PT_DFTgas}a. The pyridyl groups in T4PT support two distinct planar H-bonding motifs, which we name \textit{tip-to-tip} and \textit{head-to-tail}. In the \textit{tip-to-tip} configuration, two CH$\cdots$N interactions are present, whereas in the \textit{head-to-tail} configuration, the nitrogen atom of the pyridyl terminal group interacts with two hydrogens of the neighboring molecule, each hydrogen atom belonging to a different pyridyl group. DFT calculations at the PBE+D3 level for dimer structures in gas phase indicate a slight energetic preference for the \textit{tip-to-tip} motif. Tab.\ref{tab:bondingmotifs} summarizes the binding energies and bond lengths, which match well with prior studies on pyridine and pyridyl derivatives. \cite{Piacenza05,Sampsa2014}  \\

Additionally, T4PT can be involved in either two-fold or three-fold metal coordination bonding taking place between the N atoms of the pyridyl groups and metal atoms. In view of using Au(111) as substrate, that implies then interactions with native Au adatoms.  The two-fold Au-coordinated T4PT dimer maintains planarity (Fig.~\ref{fig_T4PT_DFTgas}d) and demonstrates significantly stronger bonding compared to the hydrogen-bonded dimers, with bond lengths comparable to previously reported values for similar metal-ligand systems.\cite{ZhangZhao2015,LiYang2012} In contrast, for the three-fold Au coordinated T4PT trimer, the pyridyl units are expected to be rotated around the C-C bond connecting the pyridyl units to the central triazine core which is caused by steric hindrance.\cite{Bischoff2016} The pyridyl rotation cannot be reproduced in fully relaxed gas-phase DFT optimizations without introducing restrictions, see Fig.~\ref{fig_T4PT_DFTgas}e. \\

On the other hand, DFT calculations for the T4PB monomer, which features a central benzene ring compared to the triazine one in T4PT, indicated that the pyridyl groups are rotated relative to the benzene core, giving a dihedral angle of 36.4$^{\circ}$ (Fig.~\ref{fig_VASPT4PB}). This leads to a non-planar molecular structure and contrasts with the fully planar one of T4PT. Notably, the metal-ligand interaction of the two-fold Au-coordinated T4PT dimer is stronger than that observed for T4PB.\\

\begin{table}[h]
    \centering
        \begin{tabular}{l p{2.0cm} cc}
        \hline\hline
        Bonding motif & Interaction energy (eV) & \multicolumn{2}{l}{Bond length (\AA)}\\
        \hline
        \multicolumn{4}{l}{T4PT}\\
         \textit{head-to-tail} & -0.17 & CH$\cdots$N &2.94 \\
         \textit{tip-to-tip} & -0.21 & CH$\cdots$N &2.56\\
         \textit{two-fold coordination}& -2.91 &N$\cdots$Au &  2.01 \\
    \multicolumn{4}{l}{T4PB}\\
\textit{two-fold coordination}& -2.38 &N$\cdots$Au & 2.01   \\
         \hline\hline
    \end{tabular}
      \caption{Interaction energies and bond lengths of various DFT-optimized (PBE+D3) T4PT and T4PB dimers in gas phase. The corresponding geometries are shown in Fig.~\ref{fig_T4PT_DFTgas} for T4PT and Fig.~\ref{fig_VASPT4PB} for T4PB.}
    \label{tab:bondingmotifs}
\end{table}

\subsection{H-bonded T4PT structures upon room temperature deposition}

\begin{figure}[t!]
	\centering 
	\includegraphics[width=0.48\textwidth]{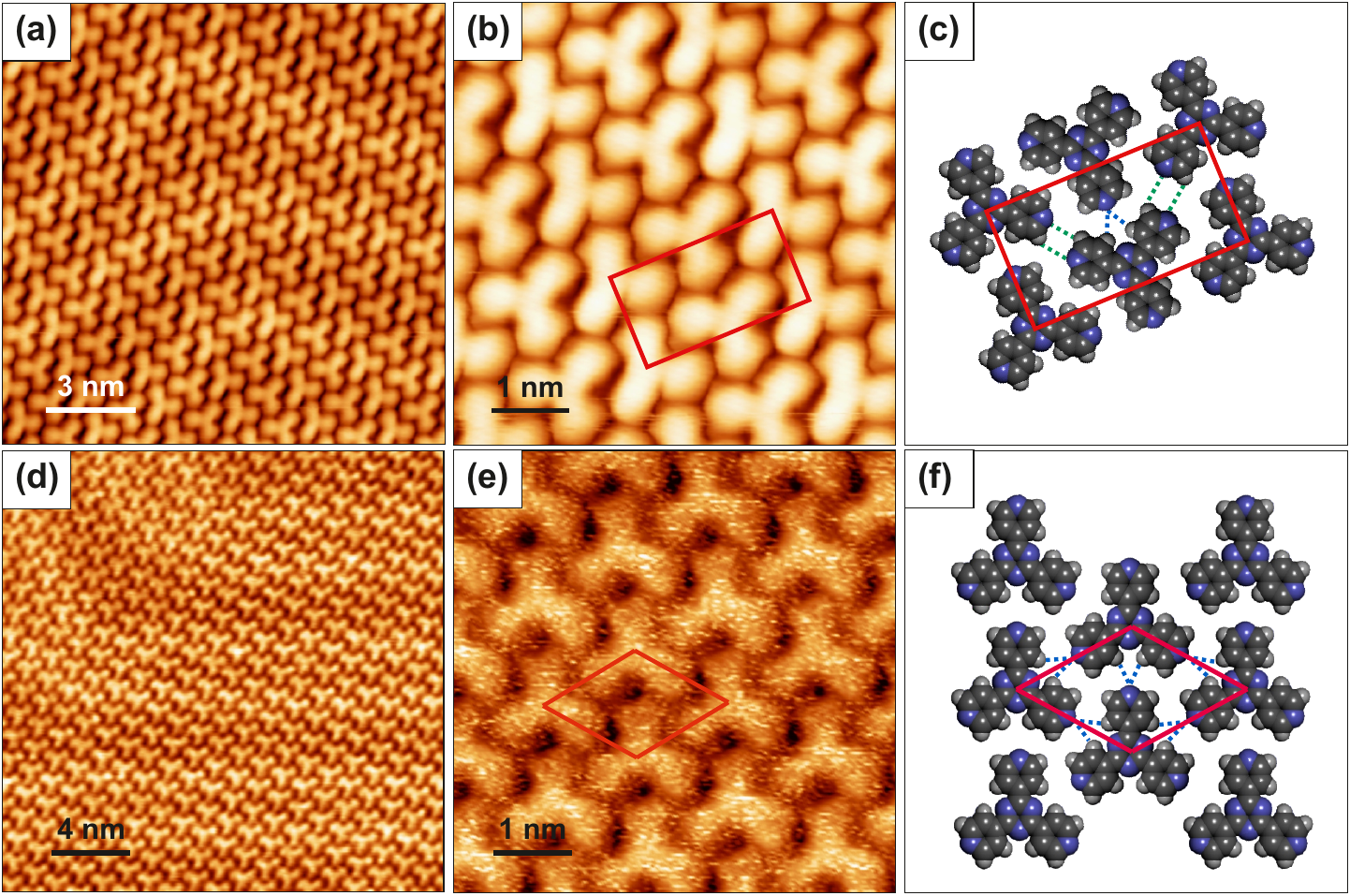}	
	\caption{(a) STM topography of the close-packed H-bonded assembly of T4PT on Au(111). (b) Detailed STM image of the close-packed H-bonded T4PT assembly. (c) Tentative model for the close-packed H-bonded assembly highlighting the \textit{head-to-tail} (blue) and the \textit{tip-to-tip} (green) H-bonds. (d) STM topography of the close-packed H-bonded assembly of T4PB on Au(111) measured at 77K. (e) Detailed STM image of the close-packed H-bonded T4PB assembly. (f) Tentative model for the close-packed H-bonded assembly highlighting the \textit{head-to-tail} (blue) H-bonds. The unit cells are indicated in red. C, N, and H atoms are depicted in gray, blue, and white, respectively. STM parameters: (a) 80~mV, 80~pA; (b) 30~mV, 80~pA; (d-e) 120~mV, 80~pA. (For interpretation of the references to color in this figure legend, the reader is referred to the web version of this article.) } 
	\label{fig_STM_RT}%
\end{figure}

Upon deposition on the Au(111) surface at room temperature, T4PT assembled in close-packed, hydrogen-bonded islands, as demonstrated by STM images shown in Fig.~\ref{fig_STM_RT}a-b. For various preparations and different coverages (up to 1 monolayer), this type of close-packed structure was consistently observed. The molecular arrangement consists of 1D molecular rows with molecules interacting via hydrogen bonds in a \textit{head-to-tail} configuration. Adjacent rows align such in alternating directions, that the molecules of neighboring rows can interact through \textit{tip-to-tip} hydrogen bonds and vdW-interactions. The resulting unit cell comprises two molecules with dimensions of $a=2.22\pm0.3$~nm, $b=1.17\pm0.3$~nm, and $\gamma=90\pm1^{\circ}$. Additionally, FFT analysis of the STM data suggests that the short side of the unit cell is rotated around $16.5\pm2^{\circ}$ relative to the Au(111) herringbone reconstruction, i.e. $<11\bar{2}>$ directions, as shown in Fig.~\ref{fig_T4PT_calibration}. However, we observe multiple rotational domains with the same unit cell dimensions. The measured unit cell of the assembly shown in Fig. \ref{fig_STM_RT} corresponds to a  $\big(\begin{smallmatrix}
  7 & 1\\
  2 & 4
\end{smallmatrix}\big)$
superstructure on Au(111). The observed structure of the hydrogen-bonded self-assembly is in line with previous studies on the 2D self-assembly of these tripod molecules \cite{Zhu2022,Lu2022} and is also consistent with Monte Carlo simulations predicting similar assemblies at close to monolayer coverage.\cite{Szabelski2016} \\

Similarly, also T4PB arranges in a close-packed H-bonded arrangement on Au(111). However, only \textit{head-to-tail} interactions were observed between the molecules, resulting in a uniform molecular orientation within an island, as shown in Fig.~\ref{fig_STM_RT}d-f.\cite{Umbach2013}  This preference may be attributed to rotated pyridyl groups in T4PB, as found by our DFT calculations and caused by steric hindrance, which renders a planar \textit{tip-to-tip} motif unfavorable.  In contrast, both bonding motifs are present for T4PT, although the \textit{head-to-tail} arrangement alone would be sufficient. This suggests that the \textit{tip-to-tip} motif provides stronger interactions for T4PT on Au(111) compared to the \textit{head-to-tail} configuration. This observation is consistent with our DFT calculations for the dimeric hydrogen-bonding motifs in the gas phase (see Tab.~\ref{tab:bondingmotifs}).

\subsection{Mixed H-bonded and metal-coordinated T4PT structures upon annealing}

\begin{figure}[t!]
	\centering 
	\includegraphics[width=0.48\textwidth]{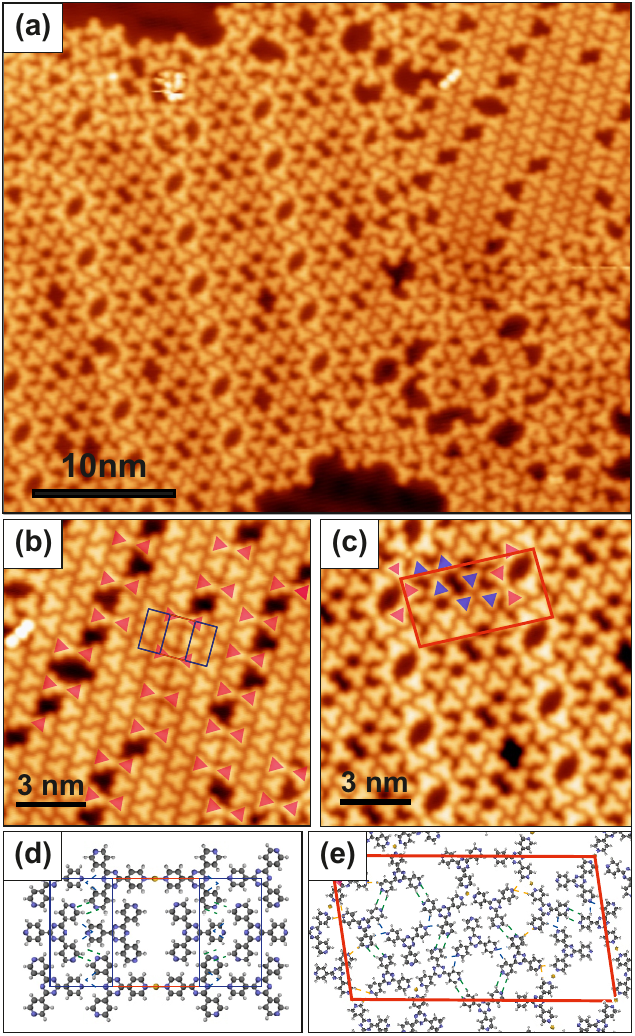}	
	\caption{(a) Overview STM image of the two observed structures upon annealing the close-packed T4PT structures at $140^{\circ}$~C. (b) Detailed STM image of the T4PT structure containing two-fold Au-coordinated dimers (red triangles) with tentative model in (d). (c) Detailed STM image of the  T4PT structure containing two-fold (red triangles) and three-fold (blue triangles) Au-coordinated units with tentative model in (e). STM parameters: (a-b-c) -10~mV, 90~pA. (For interpretation of the references to colour in this figure legend, the reader is referred to the web version of this article.)} 
	\label{fig_STM_140}%
\end{figure}

In a next step, we studied the effect of post-deposition annealing onto the molecular T4PT assembly structure. Based on earlier reports, \cite{Shi2009,Gottardi2014} the amount of native Au atoms upon temperatures larger than room temperature is sufficient to observe metal-ligand bonding to molecules having N-containing functional groups. For our case, the formation of Au-pyridyl coordination bonding should then become possible resulting in a possible formation of long-range ordered MOCNs. Importantly, the morphology of pyridyl-metal-coordinated organic structures on surfaces is highly sensitive to factors such as molecular coverage, pressure, metal-to-ligand ratio, the preferred coordination number of the metal center, and temperature.\cite{Gorbunov2023} Additionally, the rotational flexibility of the terminal pyridyl groups around the C-C bond connecting the pyridyl units to the central triazine core plays a crucial role in forming pyridyl–Au–pyridyl binding motifs, as higher coordination numbers require a rotation of the pyridyl groups to enable metal-coordination and to avoid steric hindrance preventing the binding motif.\cite{Bischoff2016} These factors are considered in the following discussion on the temperature-dependent effects on intermolecular interactions within the T4PT assemblies. \\

The representative STM image in Fig.~\ref{fig_STM_140}a shows that upon annealing at $140^\circ$C two distinct arrangements emerge on the Au(111) surface, each exhibiting different T4PT bonding motifs which are also different from what was observed at room temperature. The domain depicted in Fig.~\ref{fig_STM_140}b retains a structure similar to that observed at room temperature, as indicated by the identical unit cell highlighted in blue. In addition, few molecular vacancies and two-fold Au-coordinated dimers are present, which constitute approximately 40\% of the molecules within this domain and are highlighted in red in Fig.~\ref{fig_STM_140}b. The two-fold Au-coordinated dimers are more pronounced in STM images acquired at bias voltages around 1.3~V due to an electronic state at this energy (see Fig.~\ref{fig_T4PT_voltagedependence}). Between these two-fold Au-coordinated dimers, T4PT molecules with pyridyl units oriented towards each other are observed, as indicated by the red rectangle in Fig.~\ref{fig_STM_140}b and illustrated in the tentative model in Fig.~\ref{fig_STM_140}d. \\ 

The second domain, depicted in Fig.~\ref{fig_STM_140}c, includes hydrogen bonding as well as two- and three-fold Au-coordinated bonding motifs, resulting in a less densely packed assembly. The unit cell of this domain (outlined by the red parallelogram) contains two three-fold Au-coordinated T4PT units (blue triangles), linked via \textit{head-to-head} hydrogen bonds. At the unit cell edges, two-fold Au-coordinated trimer structures are arranged in a half-circle configuration. This domain exhibits some structural variations, as illustrated in Fig.~\ref{fig_T4PT_mixeddomainsdetail}: (i) some units are smaller in size, containing only one three-fold Au-coordinated trimer; (ii) an alternate stacking arrangement is observed for two-fold coordinated trimers at the unit cell boundary; and (iii) vacancies occur where exclusively H-bonded T4PT molecules are missing. Notably, in STM imaging, the metal atoms (typically 3d transition metals) involved in coordination bonding to the terminal pyridyl groups are typically not visible.\cite{Tait2007,Bischoff2016}\\

\begin{figure}[t!]
	\centering 
	\includegraphics[width=0.48\textwidth]{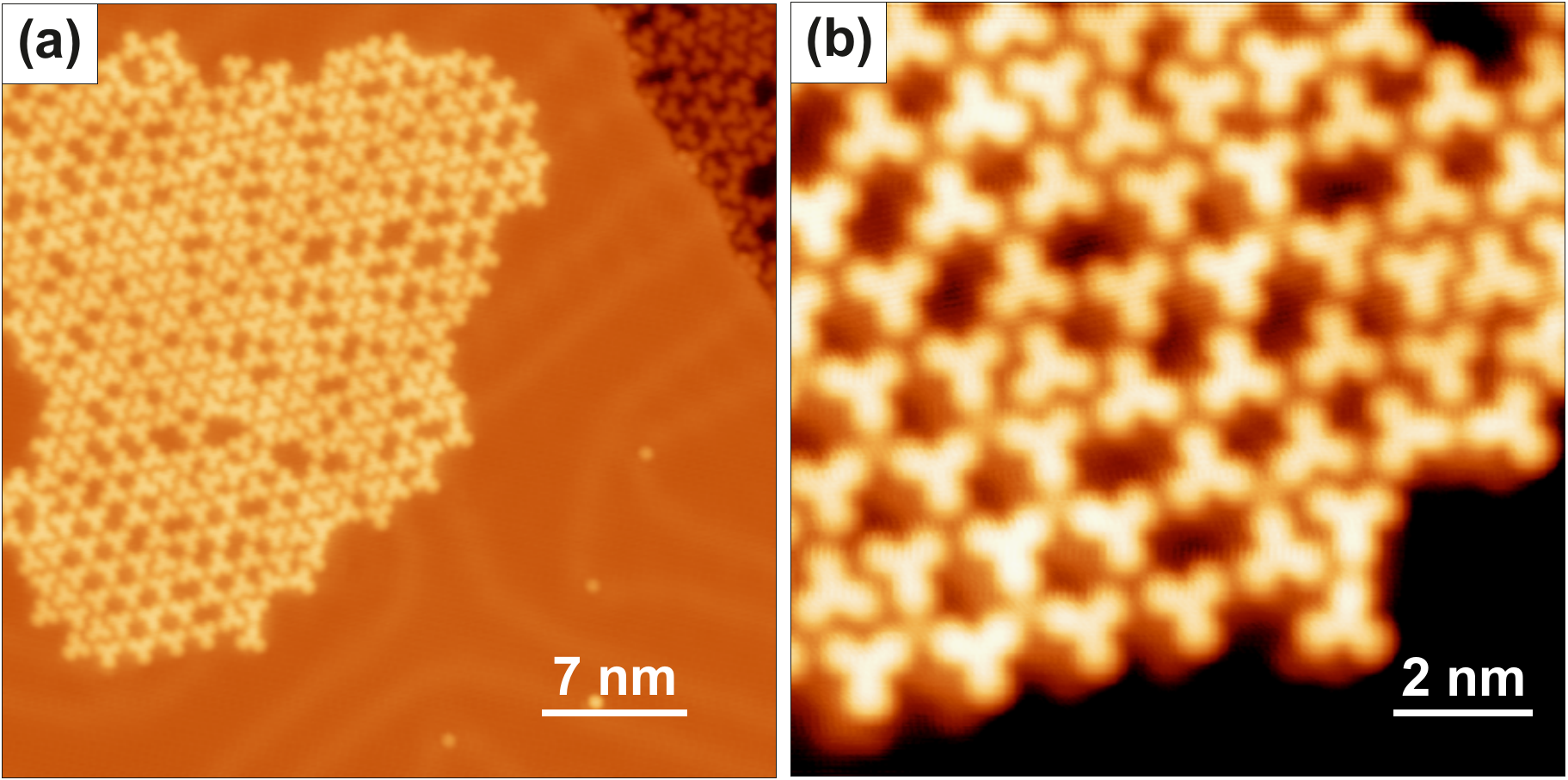}	
	\caption{(a-b) Overview and detailed STM image for T4PT submonolayer coverage deposited onto Au(111) held at $140^{\circ}$~C. In addition, to H-bonding and two-fold Au-coordination a relatively large amount of three-fold Au-coordinated molecules is visible. STM parameters: (a) 100~mV, 50~pA; (b)~100~mV, 150~pA. } 
	\label{fig_STM_140dep}%
\end{figure}

Domains composed exclusively of metal-coordinated structures were not observed for preparations at these and higher annealing temperatures; instead, the interplay between hydrogen bonding and metal–ligand interactions persists. However, when deposition was performed at elevated surface temperatures, the kinetics changed, leading to an increased number of three-fold coordinated T4PT units, as shown in Fig.~\ref{fig_STM_140dep}.\\

\begin{figure}[ht]
	\centering 
	\includegraphics[width=0.48\textwidth]{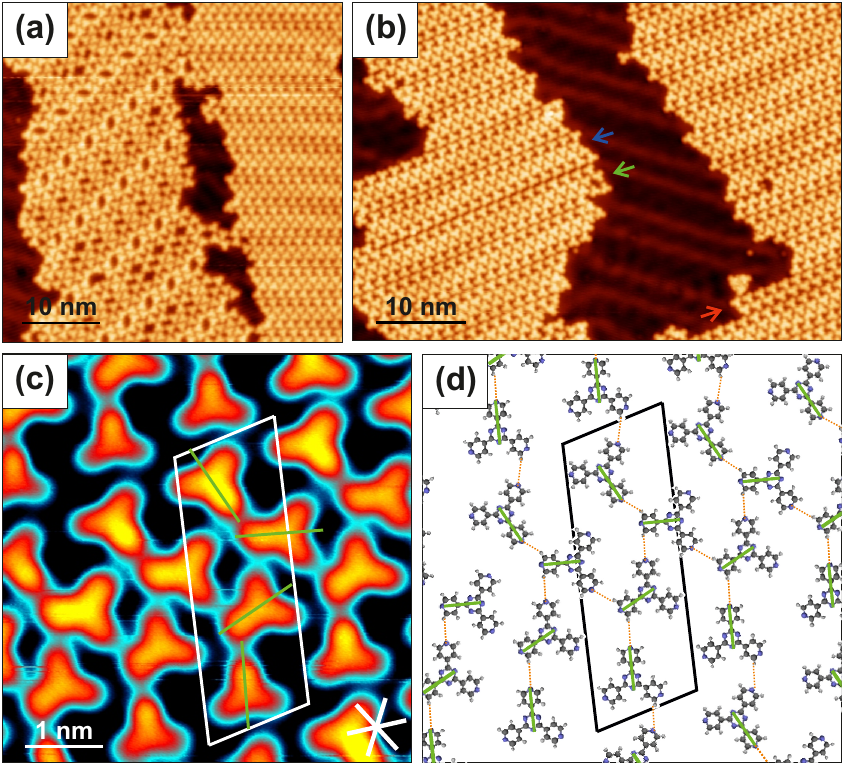}	
	\caption{(a) STM topography image after post-deposition annealing at $210^{\circ}$~C of submonolayer T4PT coverage on Au(111). Two differently ordered T4PT domains are visible: the left one matches the one shown in Fig.~\ref{fig_STM_140}c/e, while the right one exhibits a hydrogen-bonded row structure. (b) Overview STM image after post-deposition annealing at $230^{\circ}$~C. The dimeric H-bonded rows form domains with left- and right-handed configurations (blue and green arrows). The red arrow highlights an isolated dimer row. (c) Detailed STM image of the arrangement shown in (b) with the corresponding tentative model in (d). The white rhombus outlines the unit cell. The green lines indicate the mirror symmetry axes of the molecules oriented at $0^{\circ}$,$120^{\circ}$,$90^{\circ}$ and $210^{\circ}$ (from the top to the bottom in the unit cell). The white lines at the bottom right denote the principal crystallographic directions of the Au substrate. (d) Tentative model of the close-packed hydrogen bonded T4PT arrangement. The intermolecular H-bonds are highlighted by the dashed orange  ($120^{\circ}$ between pyridyls) and blue lines ($90^{\circ}$ between pyridyls), respectively. C, N, and H atoms are depicted in gray, blue, and white, respectively. STM parameters: (a) 80~mV, 80~pA; (b-c)  10~mV, 100~pA. (For interpretation of the references to color in this figure legend, the reader is referred to the web version of this article.)}
 	\label{fig_STM_210}%
\end{figure}

Upon post-deposition annealing to $210^{\circ}$~C, a new domain emerges, as shown in the overview STM images in Fig.~\ref{fig_STM_210}a-b. In this domain, the islands consist of rows that are four T4PT molecules wide (white unit cell in Fig.~\ref{fig_STM_210}c). The rows are constructed from two dimers oriented perpendicularly to each other, with T4PT molecules in the dimer oriented $120^{\circ}$ to each other. The bonding motif in the row exhibits organizational chirality, featuring both left- and right-handed domains as highlighted by the arrows in Fig.~\ref{fig_STM_210}b and Fig. \ref{fig_T4PT_chainsdetail}. The STM contrast of T4PT no longer shows three-fold symmetry, suggesting that some pyridyl rings have lost planarity with respect to the Au surface (Fig.~\ref{fig_STM_210}c). The rotation of the pyridyl unit along the C-C bond connecting the pyridyl units to the central triazine core requires thermal activation, as the planar configuration is typically stabilized by CH$\cdots$N bonds between the central triazine and the adjacent pyridyl ring. The green lines in Fig.~\ref{fig_STM_210}c indicate the primary axes of the molecules oriented at $0^{\circ}$,$120^{\circ}$,$90^{\circ}$ and $210^{\circ}$ (from top to bottom within the unit cell). The unit cell, marked by a white parallelogram, has dimensions of $a=1.43\pm 0.3$~nm and $b=3.96\pm 0.3$~nm, with an enclosing angle of $\gamma=105\pm 3^{\circ}$. Moreover, our STM measurements suggest that the short side of the unit cell is rotated approximately $10^{\circ}$ with respect to the principal Au crystallographic directions. The tentative model in Fig.~\ref{fig_STM_210}d reveals that the molecules within a row and between rows interact \textit{via} CH$\cdots$N bonds, indicating that the tilt of the pyridyl units prevents the formation of the \textit{tip-to-tip} and \textit{head-to-tail} hydrogen bonding motif. Given the observed molecule-molecule separation and that a relative orientation of 120° between the molecules in the two-fold pyridyl-coordination is rarely observed,\cite{WangYuxu2016} we do not assume the presence of coordination bonds to Au adatoms in this domain. \\

DFT calculations in gas phase of a related unit cell (see Fig.~\ref{fig_VASPT4PT120}) indicate that such a hydrogen bonded configuration is stable. The calculated interaction energy per molecule is -0.22~eV, which is sligthly stronger to that of the \textit{tip-to-tip} and \textit{head-to-tail} motif. We note, the DFT-optimized model calculated in gas phase is not directly comparable to the experimental unit cell, as the calculated unit cell is larger and consequently, most of the pyridyl units remain planar.


\subsection{Covalently bonded T4PT structures}
\begin{figure}[ht]
	\centering 
	\includegraphics[width=0.45\textwidth]{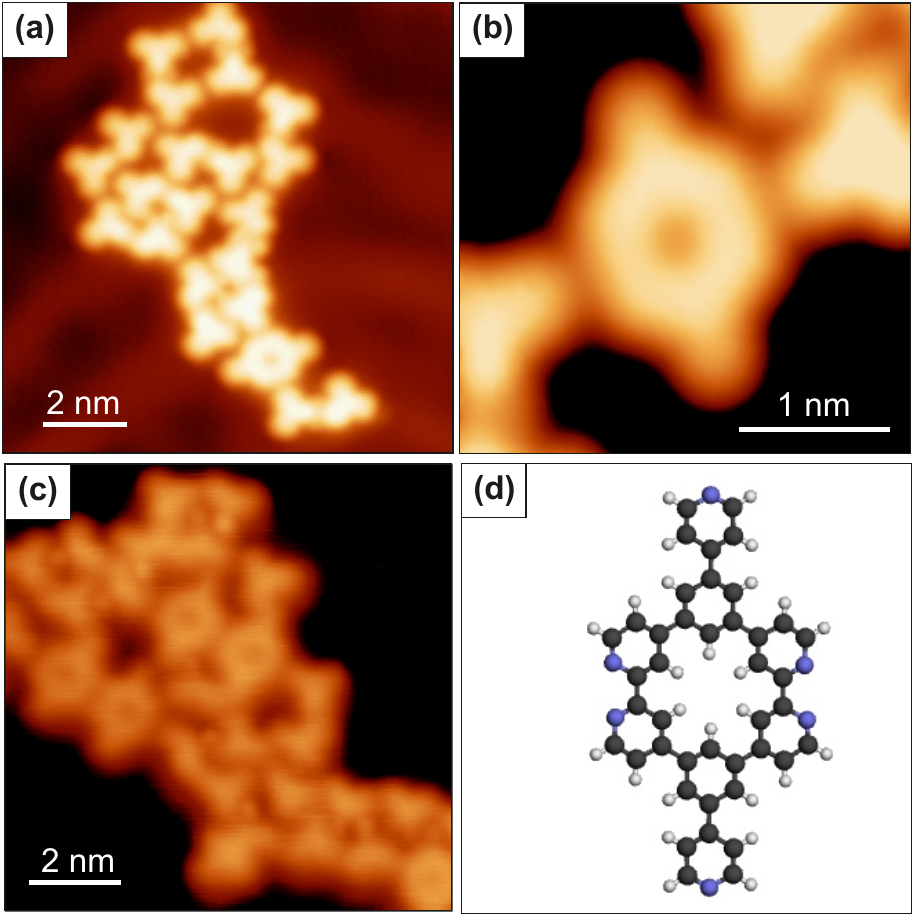}	
	\caption{(a) STM topography image taken after post-deposition annealing of T4PT submonolayer coverage on Au(111) at $230^{\circ}$~C. In addition to H-bonded and Au-coordinated molecules, a covalently-bonded dimer is visible in the lower part of the image. (b) High-resolution STM image of a covalently bonded T4PT dimer. (c) STM topography image of submonolayer coverage of T4PB on Au(111) after post-deposition annealing at $180^{\circ}$~C. Covalently linked dimers similar to the ones observed for T4PT are present. (d) Tentative structural model of the covalently linked T4PB dimer. C, N, and H atoms are depicted in gray, blue, and white, respectively.  STM parameters: (a-b) 10~mV, 100~pA; (c) -100~mV, 10~pA. (For interpretation of the references to color in this figure legend, the reader is referred to the web version of this article.)} 
	\label{fig_STM_covalent}%
\end{figure}

Upon post-deposition annealing to $230^{\circ}$~C, we observe some molecular structures with a central pore that resemble an N-substituted cyclohexa-\textit{m}-phenylene derivative, as shown in the STM images in Fig.~\ref{fig_STM_covalent}. We attribute these structures to covalently bonded T4PT dimers formed during the annealing process. The measured N-N distance between the opposing pyridyl units along the main axis of the covalently bonded dimer in Fig.~\ref{fig_STM_covalent}c is approximately 2.1~nm, in line with the distances obtained from our STM experiments. The formation of such dimers is also found for T4PB after annealing at $180^{\circ}$~C, indicating a similar covalent coupling mechanism for both molecules under thermal activation.\\ 

The covalent bonding between two T4PT molecules and between two T4TB molecules is likely activated \textit{via} a coordination-controlled \textit{ortho}-site CH-bond activation followed by dehydrogenative homocoupling.\cite{Zhang2019sm}  This reaction mechanism, previously reported on Au(111) using Fe atoms as a catalyst, is suggested by our results to proceed also with Au adatoms serving as the catalyst. Evidence for this mechanism comes from the selectivity observed in the dehydrogenative coupling: the nitrogen atoms are positioned on the same side of the newly formed C–C covalent bond, allowing them to coordinate with the same Au atom, which can also coordinate with the pyridyl group of an adjacent T4PT molecule.

\subsection{DFT calculations on the Au surface}
\begin{figure}[h!]
	\centering 
	\includegraphics[width=0.4\textwidth]{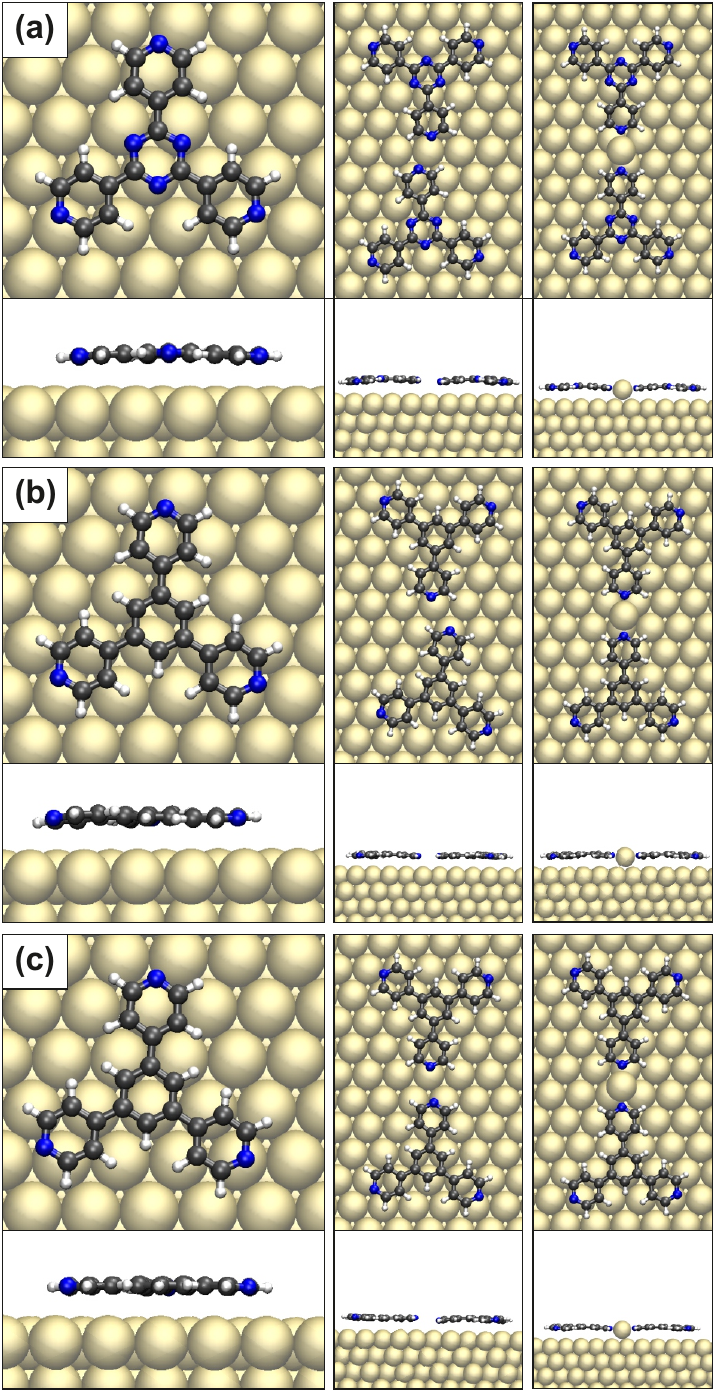}\caption{Structural models of the DFT-optimized (PBE+D3) monomer, hydrogen bonded \textit{tip-to-tip} dimer and two-fold Au-coordinated dimer (from left to right) for (a) T4PT, (b) T4PB with a non-planar configuration as starting geometry, and (c) T4PB with an initial planar configuration as starting geometry. C, N, H, and Au atoms are depicted in gray, blue, white, and yellow, respectively.  (For interpretation of the references to color in this figure legend, the reader is referred to the web version of this article.)} 
	\label{fig_STM_dftm}%
\end{figure}

To get further insight into the different interactions and to discuss differences in the adsorption geometries between T4PT and T4PB, we compare DFT calculations on the Au(111) surfaces for different monomer and dimer configurations: T4PT as well as T4PB with a planar and non-planar starting geometry. The structural models are shown in Fig.~\ref{fig_STM_dftm} and the corresponding adsorption energies are given in Tab.~\ref{tab:adsorption_energies}.

\begin{table}[h!]
\centering
\begin{tabular}{l p{1.4cm} p{1.5cm} p{1.8cm}}
\hline\hline
 & \multicolumn{3}{c}{Adsorption energy (eV) } \\ 
    & monomer & H-bonded dimer & two-fold coordination \\ \hline
\textit{T4PT}               & -2.391  &  -2.468  & -2.933 \\
\textit{T4PB planar}        & -2.583  & -2.652 &  -3.158 \\ 
\textit{T4PB non-planar}    &  -2.460 &  -2.532 & -3.021  \\ \hline\hline
\end{tabular}
\caption{Adsorption energies per T4PT/T4PB molecule for the monomer, the \textit{tip-to-tip} H-bonded and the two-fold Au-coordinated dimer on Au(111).}
\label{tab:adsorption_energies}
\end{table}

The calculations show that across all configurations --monomer, H-bonded \textit{tip-to-tip} dimer, and Au-coordinated dimer -- T4PB exhibits a stronger adsorption energy compared to T4PT.  This observation may stem from the inherently stronger adsorption of benzene relative to pyridine when both are in a flat adsorption geometry with the molecular plane parallel to the  Au(111).\cite{Ferringhi11,Tonigold00} The adsorption energy difference between T4PT and T4PB is smaller in non-planar starting configurations than the planar ones. For T4PB, a planar starting geometry consistently demonstrates stronger adsorption than the non-planar arrangement, which can be attributed to different adsorption configurations after relaxation. In the planar starting configuration of the T4PB monomer, the benzene ring exhibits an average adsorption height of approximately 330~pm above the Au surface and shows dihedral angles between the benzene plane and the pyridyl ring of 0.6–2.3°. In contrast, the non-planar configuration displays larger dihedral angles ranging from 1–11°, indicating a significant deviation from planarity. Both molecules show significant stabilization when the Au atoms coordinate with them on the Au(111) surface. However, the impact is more pronounced in T4PB, suggesting that T4PB may have more favorable electronic or geometric interactions with the Au atoms.

\section{Discussion}

Our STM study on the effect of post-deposition annealing onto the molecular T4PT assembly structures demonstrates a structural transformation from hydrogen bonded close-packed assemblies to networks with both hydrogen bonds and metal-ligand interactions between the pyridyl groups and Au adatoms. Upon annealing to $210^{\circ}$C, the molecular assemblies revert back to close-packed hydrogen bonded structures. The initial structural transition is facilitated by a rotation of the pyridyl units, enabling the formation of three-fold Au-coordinated units and leading to a more stable assembly. The reverse transformation upon annealing is attributed to the disruption of the Au-pyridyl coordination bonds. Similar reversible transitions from hydrogen bonding to coordination bonding and back have previously been observed in pyridyl-substituted porphyrins, indicating that this is not exclusive to the Au-pyridyl interaction.\cite{LiYang2012,Studener2015,Baker2021} \\

The potential interplay between hydrogen-bonding and metal-ligand interactions exists also for larger precursor molecules. However in these systems, increased structural flexibility can favor the formation of extensive hexagonal porous networks predominantly governed by metal-ligand interactions. A notable example is 1,3,5-tris[4-(pyridin-4-yl)phenyl]benzene, which incorporates an additional phenyl ring in each side arm compared to T4PB.\cite{Song2017_py} The inclusion of extra phenyl rings not only extends the molecular length but also enhances rotational freedom around the single bonds. This increased flexibility allows the molecule to adopt conformations that optimize coordination with native Au atoms on the surface, facilitating the fabrication of porous MOCN architectures.\\

\section{Summary and conclusions}

In summary, we conducted a comprehensive study of the effect of annealing on the 2D structure formation of pyridyl-functionalized tripod molecules on Au(111) at submonolayer coverages using STM. In particular, we focused on the influence of temperature on the occurrence/disappearance of specific intermolecular interactions. Our STM observations showed that T4PT forms well-ordered, H-bonded close-packed networks for coverages $\leq 1$ monolayer when deposited on Au(111) held at room temperature.  Upon annealing, these assemblies transform into more intricate ones which are in addition to H-bonding stabilized by metal-ligand bonding between the pyridyl ligands and native Au adatoms. We rationalize the kinetically-driven structure formation by unveiling the interaction strengths of the bonding motifs using DFT and compare the molecular conformation to the structurally similar pyridyl-functionalized benzene T4PB. Our findings highlight the challenge of predicting and controlling 2D structure formation, also in view of only slighly altering the molecular building block, for (porous) MOCNs on metal surfaces due to competing interactions, particularly in the case of smaller compounds.

\section*{CRediT authorship contribution statement}
\noindent\textbf{Mohammad Sajjan}: Investigation; \textbf{Neeta Bisht}: Writing – review \& editing, Investigation;  \textbf{Anjana Kanan}: Investigation; \textbf{Anne Brandmeier}: Investigation; \textbf{Christian Neiss}: Investigation; \textbf{Andreas Görling}: Writing – review \& editing, Supervision, Funding acquisition; \textbf{Meike Stöhr}: Writing – review \& editing, Conceptualization, Funding acquisition; \textbf{Sabine Maier}: Writing – review \& editing, Writing – original draft, Supervision, Investigation, Formal analysis, Conceptualization, Funding acquisition.

\section*{Declaration of competing interest}
No interest statement to declare.

\section*{Acknowledgements}
This work was funded by the German Research Foundation (DFG) through RTG 2681 (project number 491865171) at the Friedrich-Alexander-Universität Erlangen-Nürnberg and the TU Dresden and SFB 953 (project number 182849149) at the Friedrich-Alexander-Universität Erlangen-Nürnberg. Additionally, this project received funding under Horizon Europe (Energy storage in molecules, ESiM, grant agreement No 101046364). The authors gratefully acknowledge the scientific support and HPC resources provided by the Erlangen National High Performance Computing Center (NHR@FAU) of the Friedrich-Alexander-Universität Erlangen-Nürnberg (FAU) under the NHR project b146dc. NHR funding is provided by federal and Bavarian state authorities. NHR@FAU hardware is partially funded by the German Research Foundation (DFG) – 440719683.

\section*{Data availability}
Data will be made available on request.
\clearpage



\clearpage
\appendix

\section{Further STM data on the self-assembly of T4PT on Au(111)}

\begin{figure}[h]
	\centering 
	\includegraphics[width=0.48\textwidth]{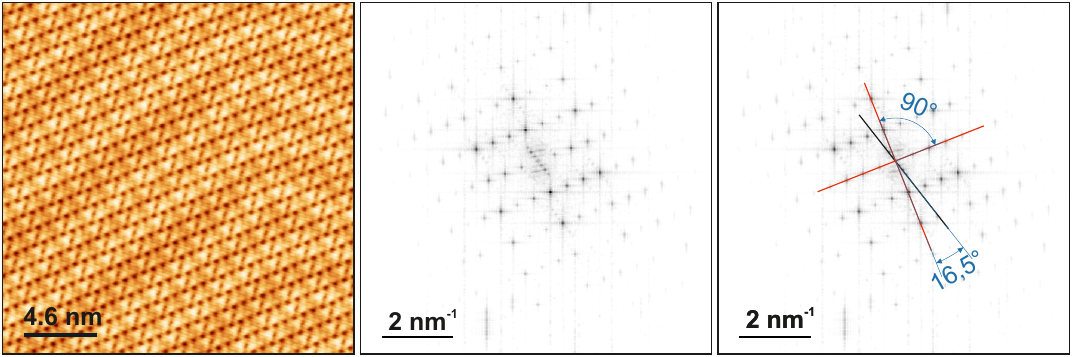}	
	\caption{\small Close-packed H-bonded assembly of T4PT upon deposition on the Au(111) surface at room temperature. (a) STM image of a large close-packed  T4PT island showing the herringbone reconstruction of Au(111) in the background (70~mV, 10~pA). (b-c) Corresponding Fast Fourier Transform (FFT) revealing the short side of the T4PT unit cell (vertical red line in c) is rotated $16.5^{\circ}\pm2^{\circ}$ relative  herringbone reconstruction of the Au(111) (black line in c). However, we observe multiple rotational domains of T4PT on Au(111), e.g. with rotation angles of $52^{\circ}\pm2^{\circ}$ and $41.4^{\circ}\pm2^{\circ}$ of the short side of the T4PT unit cell relative to the Au(111) herringbone reconstruction.} 
	\label{fig_T4PT_calibration}%
\end{figure}

\begin{figure}[h]
	\centering 
	\includegraphics[width=0.48\textwidth]{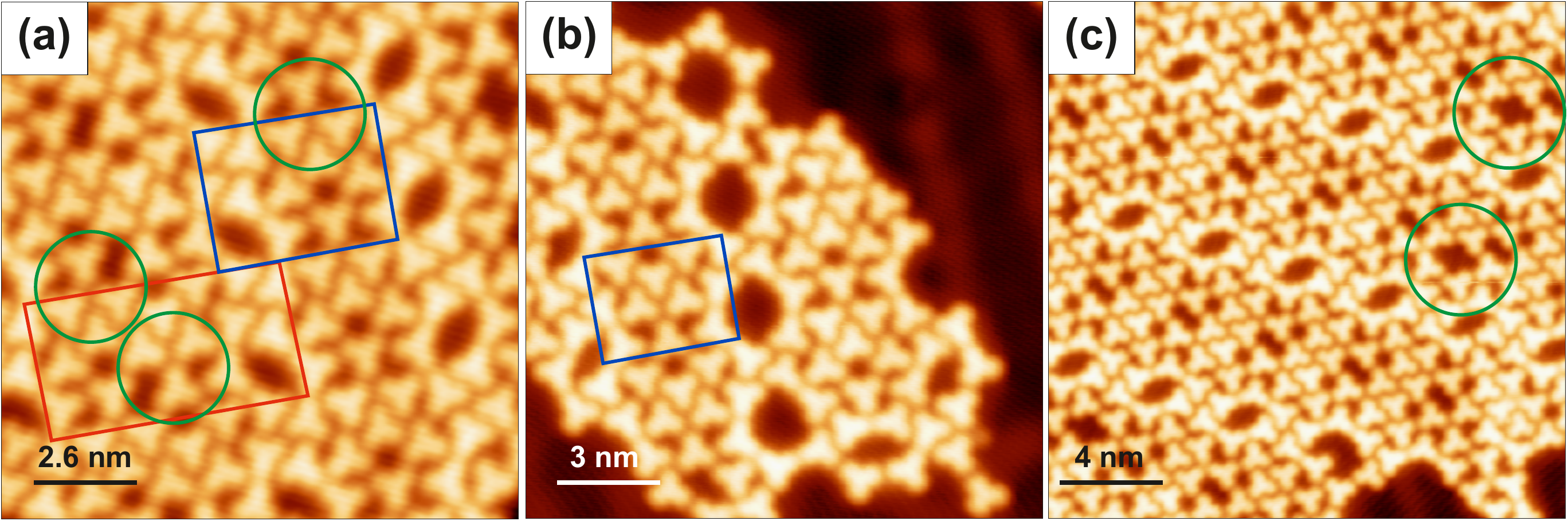}	
	\caption{\small STM topography images of T4PT structures annealed to $200^{\circ}$~C revealing detailed features of the mixed metal-coordination and H-bonding domains discussed in Fig. 3. (a) The mixed domains vary in width:  the larger unit cell (outlined in red) contains two three-fold Au-coordinated trimers (green circles), while the smaller unit cell (outlined in blue) contains one three-fold Au-coordinated trimer. (b) An alternate stacking arrangement is observed for two-fold coordinated trimers at the unit cell edge. (c) The STM image shows typical defects (green circles), which are vacancies of solely H-bonded T4PT molecules. STM parameters: (a) 20~mV, 300~pA; (b-c) 40~mV, 300~pA.} 
	\label{fig_T4PT_mixeddomainsdetail}%
\end{figure}

\begin{figure}[h]
	\centering 
	\includegraphics[width=0.48\textwidth]{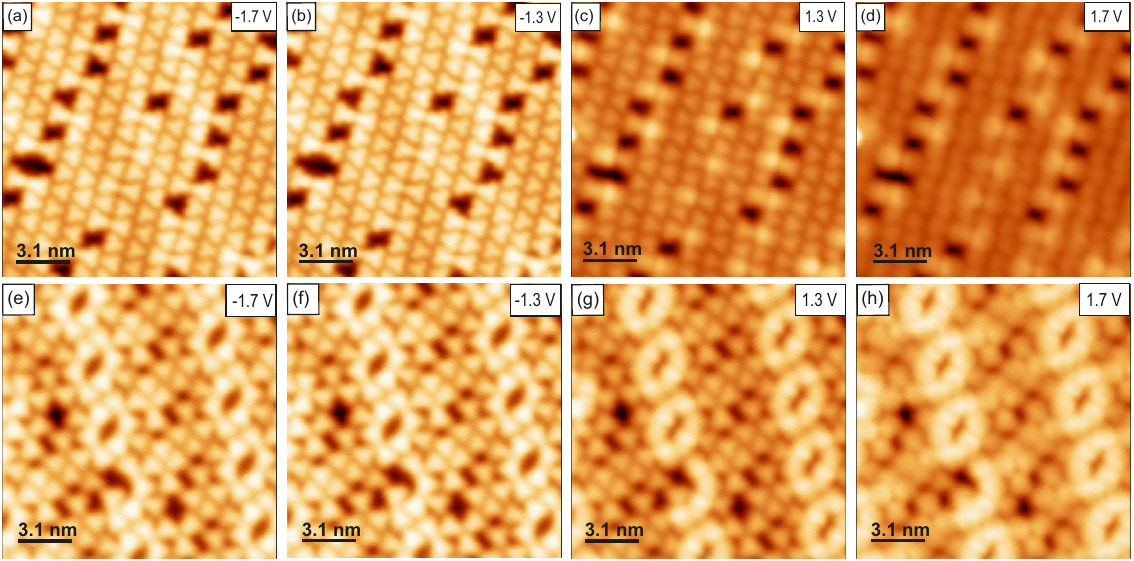}	
	\caption{\small Voltage dependent STM images of the two domains shown in Fig.~3,  illustrating the influence of bias voltage on the apparent brightness of the T4PT structures. At bias voltages around 1.3~V the two-fold Au-coordinated T4PT structures appear brighter, indicating enhanced electronic states at this energy level due to the Au coordination. } 
	\label{fig_T4PT_voltagedependence}%
\end{figure}

\begin{figure}[h]
	\centering 
	\includegraphics[width=0.48\textwidth]{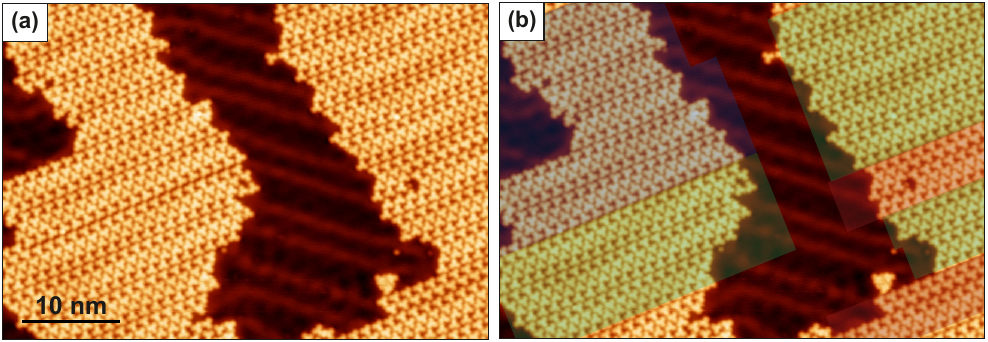}	
	\caption{\small STM topographies of the H-bonded self-assembly of T4PT upon annealing to $230^{\circ}$~C on Au(111) (10mV, 100 pA). The self-assembly exhibits organizational chirality with left- and right-handed domains, highlighted by a blue and green background. Additionally, individual dimer rows occur that are marked in red.} 
	\label{fig_T4PT_chainsdetail}%
\end{figure}


\section{Further DFT calculations}
\subsection{Gas phase calculations}
\begin{figure}[h]
	\centering 
	\includegraphics[width=0.48\textwidth]{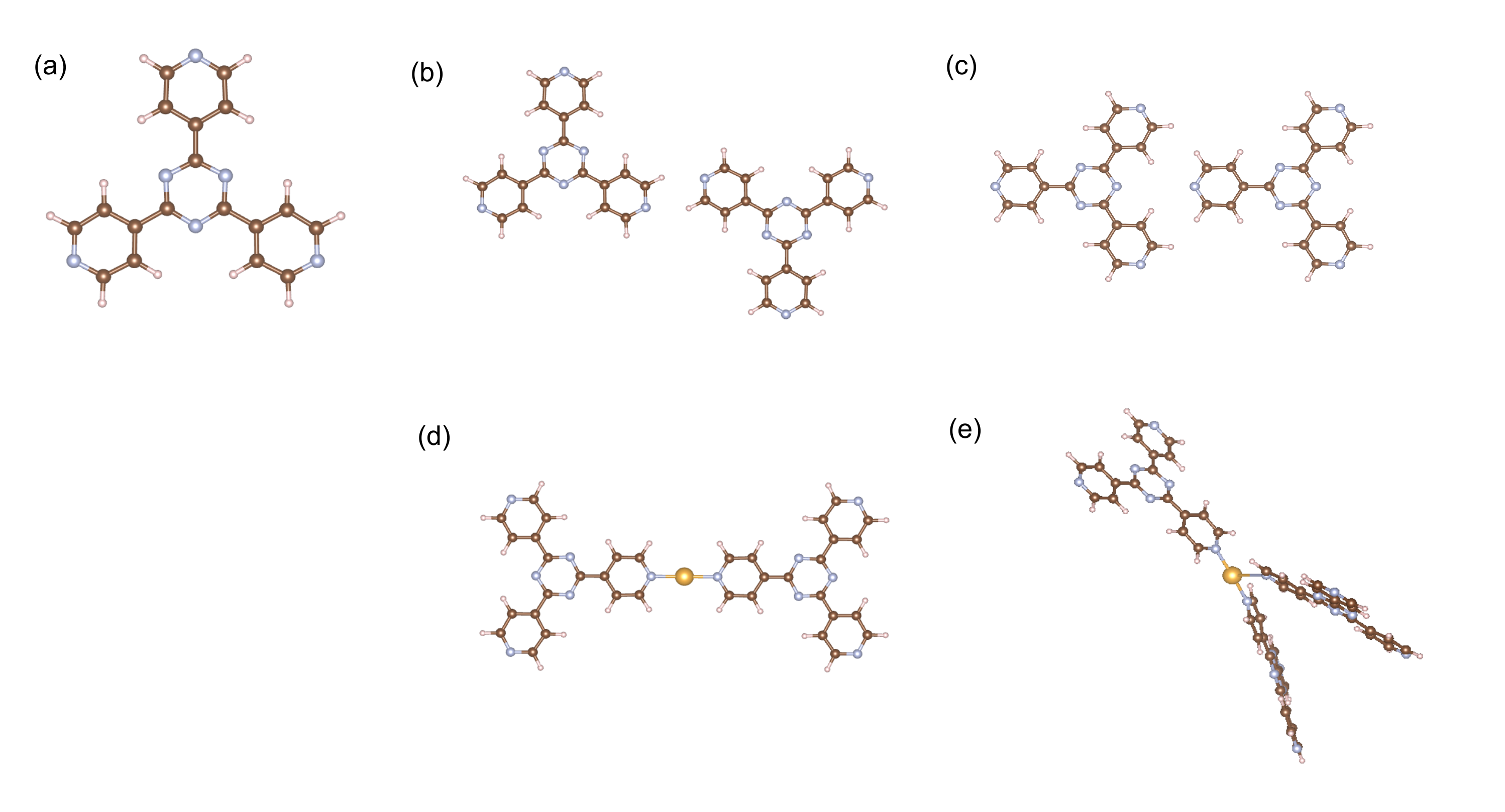}	
	\caption{\small DFT-Optimized gas phase structures of (a) T4PT monomer, (b) hydrogen-bonded \textit{tip-to-tip} T4PT dimer, (c) hydrogen-bonded \textit{head-to-tail} T4PT dimer, and (d)-(e) two-fold and three-fold Au-coordinated T4PT. C, N, H, and Au atoms are depicted in brown, blue, pink, and yellow, respectively.} 
	\label{fig_T4PT_DFTgas}%
\end{figure}

\begin{figure}[h!]
\centering
\includegraphics[width=0.48\textwidth]{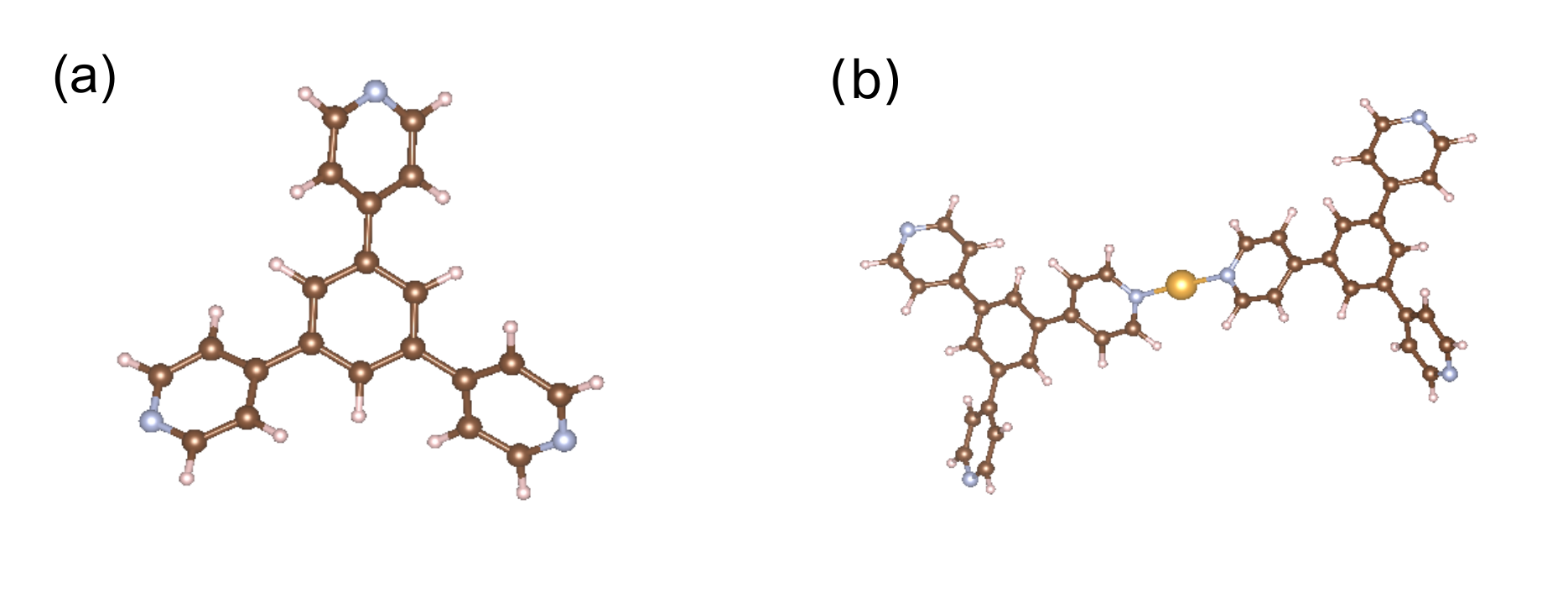}
\caption{\small DFT-Optimized gas phase structure of (a) T4PB monomer, and (b) Au-coordinated T4PB dimer. C, N, H, and Au atoms are depicted in brown, blue, pink, and yellow, respectively.}
\label{fig_VASPT4PB}
\end{figure}

\begin{table}[h!]
\centering
\begin{tabular}{lp{2.0cm}p{3.0cm}}
   \hline   \hline
\textbf{Configuration} & \textbf{Interaction Energy (eV)} & \textbf{Interaction Energy per Molecule (eV)} \\
   \hline
Fig.~S5b & \(-0.21\) & \(-0.10\) \\
Fig.~S5c & \(-0.17\) & \(-0.08\) \\
Fig.~S5d & \(-2.91\) & \(-1.45\) \\
Fig.~S5e & \(-2.96\) & \(-0.99\) \\
Fig.~S6b & \(-2.38 \) & \(-1.19 \) \\
Fig.~S7 & \(-0.87 \) & \(-0.22 \) \\
   \hline   \hline
\end{tabular}
\caption{\small Interaction energies in eV for T4PT and T4PB bonding motifs in gas-phase.}
\label{tab:interaction_energies}
\end{table}

The interaction energies to the dimeric motifs (Fig~\ref{fig_T4PT_DFTgas}b-d) are presented in Tab.~1 of the manuscript. All interaction energies for T4PT and T4PB are summarized in Tab.\ref{tab:interaction_energies}. The interaction energy \( E_{\text{int}} \) for different configurations of the T4PT system were calculated by
\begin{equation}
E_{\text{int}} = E_{\text{combined}} - \left( N_{\text{T4PT}} \times E_{\text{isolated}} + N_{\text{Au}} \times E_{\text{Au}} \right),
\end{equation}

where $E_{\text{combined}}$ is the total energy of the combined system (e.g. dimer), $E_{\text{isolated}}$ the energy of an isolated T4PT molecule, $E_{\text{Au}}$  the energy of a single Au atom (used in two-fold and three-fold coordination), $N_{\text{T4PT}}$ the number of T4PT molecules in the system, and $N_{\text{Au}}$ the number of Au atoms in the system.

\begin{figure}[h!]
\centering
\includegraphics[width=0.4\textwidth]{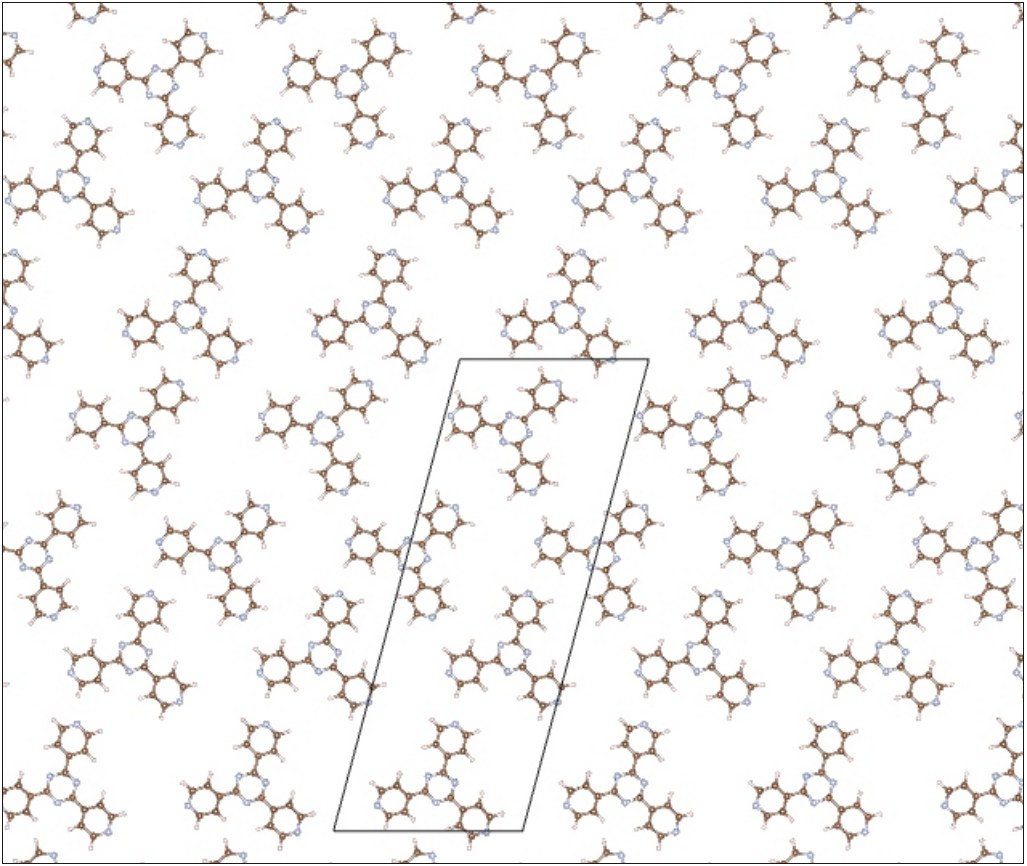}
\caption{\small Optimized gas phase of structure of hydrogen bonded T4PT structure. The unit cell is highlighted by the black parallelogram and has a dimension of $a=16.36$~\AA,  $b=42.25$~\AA,  and $\gamma=75^{\circ}$. C, N, H, and Au atoms are depicted in brown, blue, pink, and yellow, respectively.}
\label{fig_VASPT4PT120}
\end{figure}

\end{document}